\documentclass{aa}  

\usepackage{graphicx}
\usepackage{txfonts}
\usepackage{natbib} 
\bibpunct{(}{)}{;}{a}{}{,} 

\usepackage{verbatim} 
\usepackage{hyperref}   
\hypersetup{colorlinks=true,linkcolor=blue,citecolor=blue,filecolor=blue,urlcolor=blue}

\usepackage{makeidx}
\makeindex

\begin{document}

   \title{Testing the asteroseismic estimates of stellar radii with surface brightness-colour relations and {\it Gaia} DR3 parallaxes}  

   \subtitle{Red giants and red clump stars}
  \author{G. Valle \inst{1, 2}, M. Dell'Omodarme \inst{1}, P.G. Prada Moroni
        \inst{1,2}, S. Degl'Innocenti \inst{1,2} 
}
\titlerunning{APO-k2 teo}
\authorrunning{Valle, G. et al.}

\institute{
        Dipartimento di Fisica "Enrico Fermi'',
        Universit\`a di Pisa, Largo Pontecorvo 3, I-56127, Pisa, Italy
        \and
        INFN,
        Sezione di Pisa, Largo Pontecorvo 3, I-56127, Pisa, Italy
}

   \offprints{G. Valle, valle@df.unipi.it}

   \date{Received ; accepted 09/09/2024}

  \abstract
{}
{
We compared stellar radii derived from asteroseismic scaling relations with those estimated using two independent surface brightness-colour relations (SBCRs) combined with {\it Gaia} DR3 parallaxes. 
}
{
We cross-matched asteroseismic and astrometric data for over 6,400 red giant branch (RGB) and red clump (RC) stars from the APO-K2 catalogue with the TESS Input Catalogue v8.2 to obtain precise $V$ band magnitudes and $E(B-V)$ colour excesses. 
We then adopted two different SBCRs from the literature to derive stellar radius estimates, denoted as $R^a$ and $R^b$, respectively. 
We analysed the ratio of these SBCR-derived radii to the asteroseismic radius estimates, $R,$ provided in the APO-K2 catalogue.
}
{
Both SBCRs exhibited good agreement with asteroseismic radius estimates. On average, $R^a$ was overestimated by 1.2\% with respect to $R$, while $R^b$ was underestimated by 2.5\%. For stars larger than 20 $R_{\sun}$, SBCR radii are systematically lower than asteroseismic ones.
The dispersion in the radius ratio was similar for the two methods (around 10\%). 

The agreement with asteroseismic radii shows a strong dependence on the parallax. The dispersion is halved for stars with a parallax greater than 2.5 mas. In this subsample, $R^b$ showed perfect agreement with $R$, while $R^a$ remained slightly overestimated, by 3\%.
A trend with [Fe/H] 
was found at a level of 4\% to 6\% per dex.
Additionally, a clear trend with asteroseismic mass is found. For stars less massive than about 0.95 $M_{\sun}$,  SBCR radii were significantly higher than asteroseismic ones, by about 6\%. This overestimation correlated with the presence of extended helium cores in these stars' structures relative to their envelopes.
Furthermore, radius ratios showed a dichotomous behaviour at higher masses, mainly due to the presence of several RC stars with SBCR radii significantly lower
with respect to asteroseismology. This behaviour
originates from a different response of asteroseismic scaling relations and SBCR to [$\alpha$/Fe] abundance ratios for massive stars, both in RGB and RC phases, which is reported here for the first time.}
{} 

   \maketitle

\section{Introduction}\label{sec:intro}

Accurate measurements of stellar masses and radii are crucial to constraining stellar structure and evolution models. 
High-quality measurements of these parameters can be 
obtained from detached double-lined eclipsing binaries, so these systems are routinely adopted to test stellar evolution models and to constrain their free parameters \citep[see among many][]{Andersen1991b, Torres2010,  TZFor, Claret2017, AiPhe}.

Obtaining precise estimates of masses and radii from single stars is more problematic.
Spectroscopic determinations of the effective temperature, surface gravity,
and metallicity of a star, combined with a known parallax, can provide the stellar radius and luminosity. However, the
strong dependence of the spectroscopic atmospheric parameters on the
atmosphere models and on the method used for their derivation hamper firm predictions \citep[see e.g.][]{Torres2012, Coelho2014,  Ivanyuk2017, Branco2024}. As a matter of fact, differences of about 60 K in effective temperature and 0.15 dex in $\log g$
exist between determinations from different surveys for red giant branch (RGB) stars \citep{Hegedus2023, Yu2023}.

A different approach to evaluating the linear radius of a star is to combine an estimate of its angular radius with its distance, which has been made particularly attractive by the recent determination of accurate distances for over a billion stars from {\it Gaia} Data Release 3 \citep[DR3;][]{Gaia2021}.
Very accurate angular size measurements can be obtained with long-baseline interferometry \citep[LBI; e.g.][]{
Nordgren1999, Mozurkewich2003, Baines2010, Gallenne2012, Lachaume2019, Perraut2020}. However, LBI is limited to stars with angular diameters greater than about 0.5 mas \citep{Gallenne2018}, rendering it impractical for most targets.
Surface brightness–colour relations (SBCRs) provide an efficient alternative for determining stellar angular diameters from photometric measurements. Essentially, SBCRs establish a link between a star's angular size and its de-reddened brightness across various photometric bands. SBCRs are usually
calibrated using samples of stars with well-determined LBI radii \citep[see e.g.][]{Kervella2004,  DiBenedetto2005, Salsi2021}.
More than 20 SBCRs exist in the literature, and many of them focus on the $V$ and $K$ bands, as this set of colours provides the lowest dispersion \citep{Kervella2004}.
Comparisons among recent SBCRs have revealed a limited variability for late-type stars \citep[e.g.][]{Pietrzynski2019, Salsi2022, Nardetto2023}. For instance, the comparison among 19 SBCRs performed by \citet{Nardetto2023} showed an agreement better than about 0.008 mag between 1.5 and 2.5 mag in $V - K$. The disagreement increased outside this range.
SBCRs play a fundamental role in the distance determination of eclipsing binaries.
As an example, a precise SBCR calibrated on 41 nearby red clump (RC) giant stars \citep{Gallenne2018} has allowed the distance to the Large Magellanic Cloud to be estimated with a precision of 1\% \citep{Pietrzynski2019}. 

A new method for estimating stellar radii has emerged in recent years. The
growth of observational asteroseismology, thanks to satellite missions 
such as \textit{Kepler} and Transiting Exoplanet Survey Satellite \citep[TESS;][]{Borucki2010, Ricker2015},
has opened a new way to estimate stellar properties of stars, such as mass, radius, and age. These data allow the RGB stellar radii to be estimated with a precision of about 4\% 
\citep{Pinsonneault2014, Martig2015, Valle2024-alpha}. 
Notably, neither the distance from the observer nor the reddening influence the asteroseismic estimates.
In fact, this method requires global seismic parameters, namely, the frequency
of maximum power, $\nu_{\rm max}$, a large frequency separation, $\Delta \nu$, and a determination of the stellar effective temperature. 
Given these ingredients, scaling relations yield the asteroseismic estimates of mass and radius  \citep{Ulrich1986, Kjeldsen1995}. While the validity of these scaling relations in the RGB phase has been questioned \citep[e.g.][]{Epstein2014, Gaulme2016, Viani2017, Brogaard2018, Buldgen2019}, corrections accounting for the temperature and metallicity of the star have been proposed in the literature \citep[e.g.][]{Zinn2022, Stello2022}.
For stars with a metallicity [Fe/H]$ > -1.0$, scaling relations are considered reliable \citep{Epstein2014, Valentini2019, Stasik2024}. 
In the following, we adopt asteroseismic radii as a reference. It is, however, important to clarify that we do not assume them to be the gold standard for radius determination. Disagreement in the measurement of radii from asteroseismic-corrected scaling relations and interferometric measurements have been reported by different authors, while other researchers have not found significant differences \citep[see e.g.][and references therein]{Hekker2020, Thomsen2022}.

Given the relevant differences between asteroseismic and SBCR methods, a comparison of their results obtained when applied to a homogeneous sample is of particular interest. 
The recently released APO-K2 catalogue \citep{Stasik2024} offers a great opportunity to perform this test, as this catalogue contains 
high-precision data of more than 7,500 RGB and RC stars, combining  spectroscopic \citep[APOGEE DR17;][]{Abdurrouf2022}, asteroseismic \citep[K2-GAP;][]{Stello2015}, and astrometric \citep[\textit{Gaia} DR3;][]{Gaia2021} data.

\section{Adopted SBCRs and data selection}

The surface brightness,  $S_{\lambda}$, of a star is linked to its limb-darkened angular diameter, $\theta,$ and its apparent magnitude
corrected from the extinction, $m_{\lambda0}$. In the $V$ band, $S_V$ is defined as
\begin{equation}
 S_V = V_0 + 5 \log \theta,   \label{eq:sv1}
\end{equation}
where $V_0$ is the $V$ band magnitude corrected for extinction. From Eq.~(\ref{eq:sv1}) it follows that
\begin{equation}
\theta = 10^{0.2 \; (S_V - V_0)}. \label{eq:theta}
\end{equation}
Therefore, an estimate of the stellar linear radius is
\begin{equation}
    r = 0.5 \; d \; \theta, \label{eq:r}
\end{equation}
where $d$ is the heliocentric distance of the star.

We adopted two different SBCRs. The first, proposed by \citet{Pietrzynski2019}, is
\begin{equation}
S_V^a = 1.330  [(V - K)_0 - 2.405] + 5.869 \; {\rm mag},
\end{equation}
where $(V-K)_0$ is the colour corrected for reddening. This relation was fitted in a colour range $(V - K)_0$ from 2.0 to 2.8 mag.
The second SBCR we adopted for testing was proposed by \citet{Salsi2021}. They used a slightly different formalism in their paper, but it can be written as 
\begin{equation}
S_V^b = 1.22 (V - K)_0 + 2.864 \; {\rm mag}.
\end{equation}
This relation (from Table 5 in \citealt{Salsi2021} for F5/K7-II/III spectral class stars) is valid in the range $(V - K)_0$ from 1.8 to 3.9 mag.

The APO-K2 catalogue provides the $K_s$ band magnitude from Two Micron All-Sky Survey 
and the parallax from {\it Gaia} DR3, corrected according to the {\it Gaia} zero-point  \citep{Lindegren2021}.
However, the $V$ band magnitudes it provides are computed start\index{start}ing from magnitudes in the $J$ and $K_s$ bands \citep{Stasik2024} and have low precision. The APO-K2 catalogue does not provide information about colour excess $E(B-V)$ or extinction in the $V$ band $A_V$ either. 
To obtain both more precise $V$ band magnitudes and $E(B-V)$, we cross-matched the APO-K2 catalogue with the TESS Input Catalogue (TIC) v8.2. The TIC adopts the three-dimensional empirical dust maps from Panoramic Survey Telescope and Rapid Response System \citep{Green2018}, with a re-calibration coefficient of 0.884 applied to obtain  $E(B-V)$ values, as prescribed by \citet{Schlafly2011}.
Nevertheless, it is known that the extinction has a negligible effect on the SBCR because the colour and surface brightness are sensitive to it in almost the same way. The only (tiny) difference comes from the extinction in the $K$ band \citep{Nardetto2020}. We adopted the following extinction relations from \citet{Cardelli1989}: $A_V = 3.1 \, E(B-V)$ and $A_K = 0.114 \, A_V$. We also verified that the adoption of the relation $A_K = 0.089 \, A_V$ from \citet{Nishiyama2009} does not modify the results; it leads to an average variation in the estimated radii of only 0.2\%. 

Data in the APO-K2 catalogue were subjected to a selection procedure, both to reject apparent outliers and to restrict the data to a metallicity range where asteroseismic scaling relations are the most reliable. Stars satisfying all the following constraints were retained in the sample: [Fe/H] > $-1.0$ dex, 4000 K < $T_{\rm eff}$ < 5300 K, $-0.1$ dex < [$\alpha$/Fe] < 0.4 dex, and $\log g$ < 3.25. Stars in the RGB with a mass lower than 0.75 $M_{\sun}$ were rejected as artefacts because single stars with such a low mass cannot be in the RGB given their long evolutionary timescale. Stars outside the range of colour for the \citet{Salsi2021} SBCR were excluded and so were stars with a binary flag set in the APO-K2 catalogue. Finally, stars with a relative error in the parallax greater than 0.1 were rejected. This allowed us to rely on distances obtained using the inverse parallax \citep{Bailer2021, Fouesneau2023}.  
The final sample comprises 6,420 stars, with 4,202 of them in the RGB and 2,218 in the RC phase.

\section{SBCRs to asteroseismic radii comparisons}\label{sec:results}

\begin{figure}
        \centering
        \resizebox{\hsize}{!}{\includegraphics{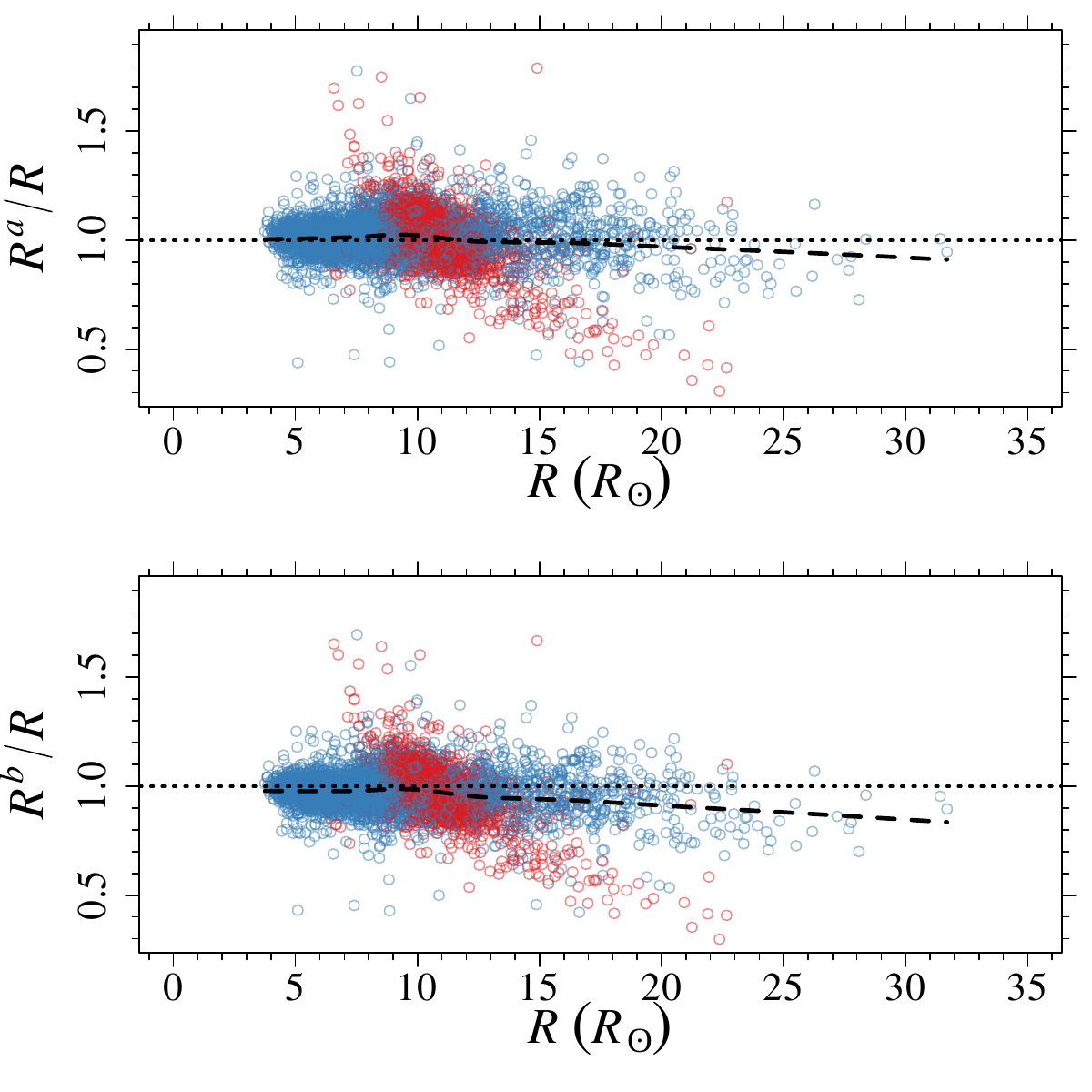}}
        \caption{Ratio of the radii from SBCRs and from asteroseismology.   {\it Top}: Radius ratio using the \citet{Pietrzynski2019} SBCR. Red and blue points correspond to RC and RGB stars, respectively. The dashed black line is a smoother of data, while the dotted one serves as a visual aid. {\it Bottom}: Same as in the top panel but using the \citet{Salsi2021} SBCR.  }
        \label{fig:R-R2-resid}
\end{figure}

For all stars in the final sample, linear radii were obtained from 
Eq.~(\ref{eq:r}), with $\theta$ (Eq.~(\ref{eq:theta})) computed using
the SBCR from \citet{Pietrzynski2019} and \citet{Salsi2021}. This resulted in estimates denoted as $R^a$ and $R^b$, respectively.
The comparison of these radii with $R$ from asteroseismic scaling relations (Fig.~\ref{fig:R-R2-resid}) revealed both similarities and differences. 

Overall, there is a notable agreement between the two SBCR methods and the predictions from the scaling relations.
The agreement between the SBCR methods and the asteroseismic radius, $R,$ is better for $R^a$, with an overestimation of the radii by 1.2\% over the full sample, while $R^b$ leads to an underestimation by 2.5\%.
Both SBCRs underestimate radii at higher values, a trend more pronounced for $R^b$ compared to $R^a$. 
Interestingly, the analysed SBCRs show the same feature for RC stars because they strongly underestimate radii larger than about 12 $R_{\sun}$ for RC stars, while the trend for RGB stars is much less pronounced. This difference, which we discuss in detail later, is likely due to differences in the mass and metallicity distribution between the samples of RC and RGB stars. Finally, the dispersion is similar for the two SBCRs, with a standard deviation of 10\%.

\begin{figure}
        \centering
        \resizebox{\hsize}{!}{\includegraphics{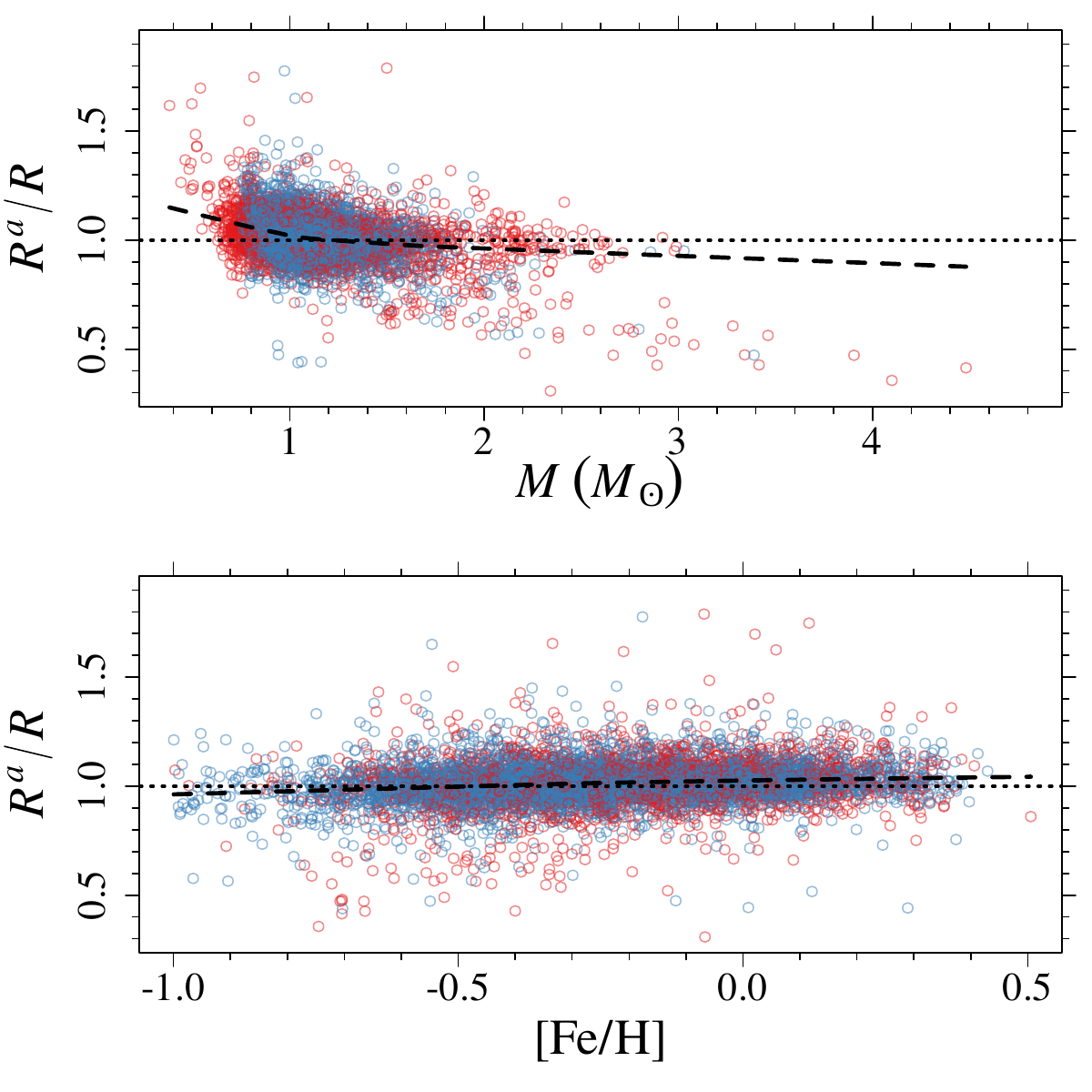}}
        \caption{Ratio of the radii from the \citet{Pietrzynski2019} SBCR and from asteroseismology as a function of different stellar parameters.  {\it Top}: Radius ratio as a function of the stellar mass as estimated by scaling relations. {\it Bottom}: Same as in the top panel but as a function of the metallicity [Fe/H]. Colour codes and line styles are as in Fig.~\ref{fig:R-R2-resid}.   }
        \label{fig:R-mass-feh}
\end{figure}

Figure~\ref{fig:R-mass-feh} explores the dependence of radius ratios on various stellar parameters, focusing on the results obtained using the SBCR from 
\citet{Pietrzynski2019}. Similar trends were obtained for the \citet{Salsi2021} SBCR. 
The impact of the metallicity is investigated in the bottom panel of Figure~\ref{fig:R-mass-feh}. There is a  weak positive trend of $R^a/R$ with metallicity [Fe/H]. This trend is approximately 6\% per dex for $R^a/R$, and it is about 4\% per dex for $R^b/R$.

\begin{figure}
        \centering
        \resizebox{\hsize}{!}{\includegraphics{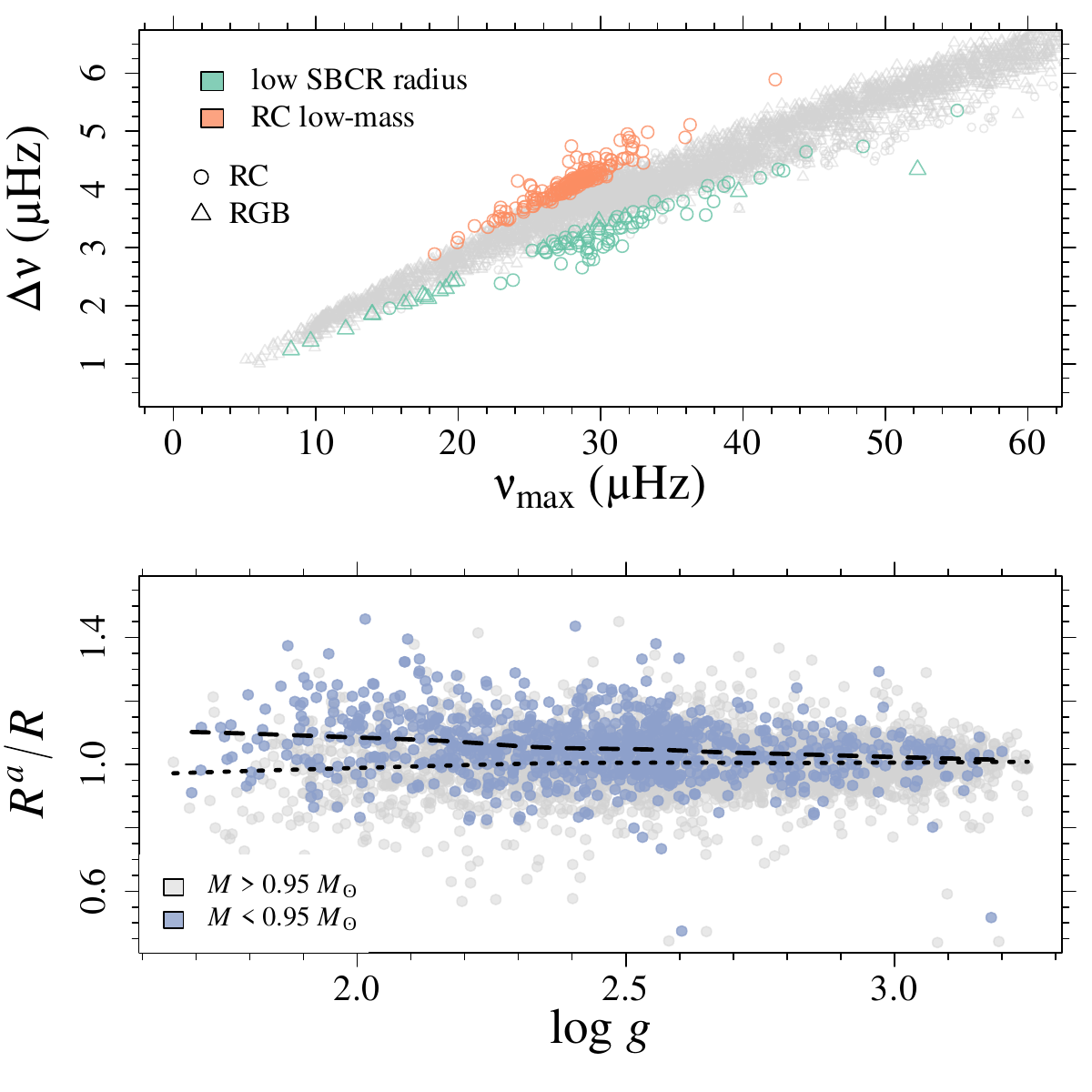}}
        \caption{Seismic parameters distribution and dependence on $\log g$ for the $R^a/R$ ratio. {\it Top row}: Global seismic parameters for the investigated sample. Green symbols identify stars with $R^a/R < 0.85$. Orange symbols identify RC stars with a mass lower than 0.75 $M_{\sun}$.  {\it Bottom row}: Dependence on $\log g$ for RGB stars in different mass bins. The dotted and dashed lines are a smoother of the sample with $M > 0.95~M_{\sun}$ and  $M < 0.95~M_{\sun}$, respectively.}
        \label{fig:dni-nimax-rgb}
\end{figure}

A significant dependence on asteroseismic mass can be observed in the top panel of Fig.~\ref{fig:R-mass-feh}. At lower masses, the SBCR method clearly provides a larger radius. However, in the mass range 1.0 to 2.0 $M_{\sun}$, the average discrepancy is less than 0.1\%. The difference at the lower mass end is partially due to the presence of 189 RC stars with masses below 0.75 $M_{\sun}$, whose SBCR radii are larger than asteroseismic ones by a median of about 6\%. Excluding RC stars less massive than 0.75 $M_{\sun}$, as done for the RGB sample, is however not theoretically justified. In fact, the mass loss, occurring primarily during the later stages of the RGB phase, may significantly impact the stellar mass at the clump phase. For instance, studies of globular clusters suggest that progenitor stars with initial masses of 0.80 $M_{\sun}$ can lose around 0.17 $M_{\sun}$ during their evolution \citep{Howell2022}. 
The internal structure of these objects is dominated by a helium core that is roughly 0.5 $M_{\sun}$ in size, with minimal variation depending on the star's initial chemical composition or mass \citep[see e.g.][]{Sweigart1978, Cassisi2016}. Asteroseismic observations, which are sensitive to a star's internal structure, can readily identify these stars, as shown in the top panel of Fig.~\ref{fig:dni-nimax-rgb}. 
These low-mass RC stars exhibit a higher $\Delta \nu$ value at fixed $\nu_{\rm max}$ with respect to more massive objects.
However, the presence of these low-mass RC stars alone does not fully explain the discrepancy between SBCRs and asteroseismic radii. This is because even for RGB stars below 0.95 $M_{\sun}$,  $R^a/R$ is overestimated by a median of 4\%. 
This overestimation is likely due to the significant helium core that grows steadily as the stars ascend the RGB. 
The bottom panel of Fig.~\ref{fig:dni-nimax-rgb} demonstrates a clear dependence of the $R^a/R$ ratio on the stellar $\log g$. Stars with lower $\log g$ values have a larger helium core, which in turn leads to a greater bias in the $R^a/R$ measurement.
A possible reason for the detected discrepancies is the difference in mass range used for calibration. For example, the \citet{Gallenne2018} sample used to calibrate the \citet{Pietrzynski2019} SBCR relation did not include any stars with an estimated mass below 0.9 $M_{\sun}$. It is, however, interesting that the theoretical investigation performed by \citet{Salsi2022} reports negligible differences between $\log g = 3$ and $\log g = 0$ in the predicted SBCR radii. Therefore the trend detected here might also be due to an inherent bias in the asteroseismic relations.  

For higher masses, dominated by RC stars, a more complex and dichotomous trend emerges. Some SBCR radii are estimated close to the asteroseismic ones, while others have significantly lower values. The latter group corresponds to the lower end of the arc-shaped distribution for RC stars in Figure~\ref{fig:R-R2-resid}. The top panel in Figure~\ref{fig:dni-nimax-rgb} shows that objects with $R^a/R < 0.85$ correspond to the lower end of the $\Delta \nu$ - $\nu_{\max}$ distribution.

\begin{figure*}
\centering
         \includegraphics[width=17cm]{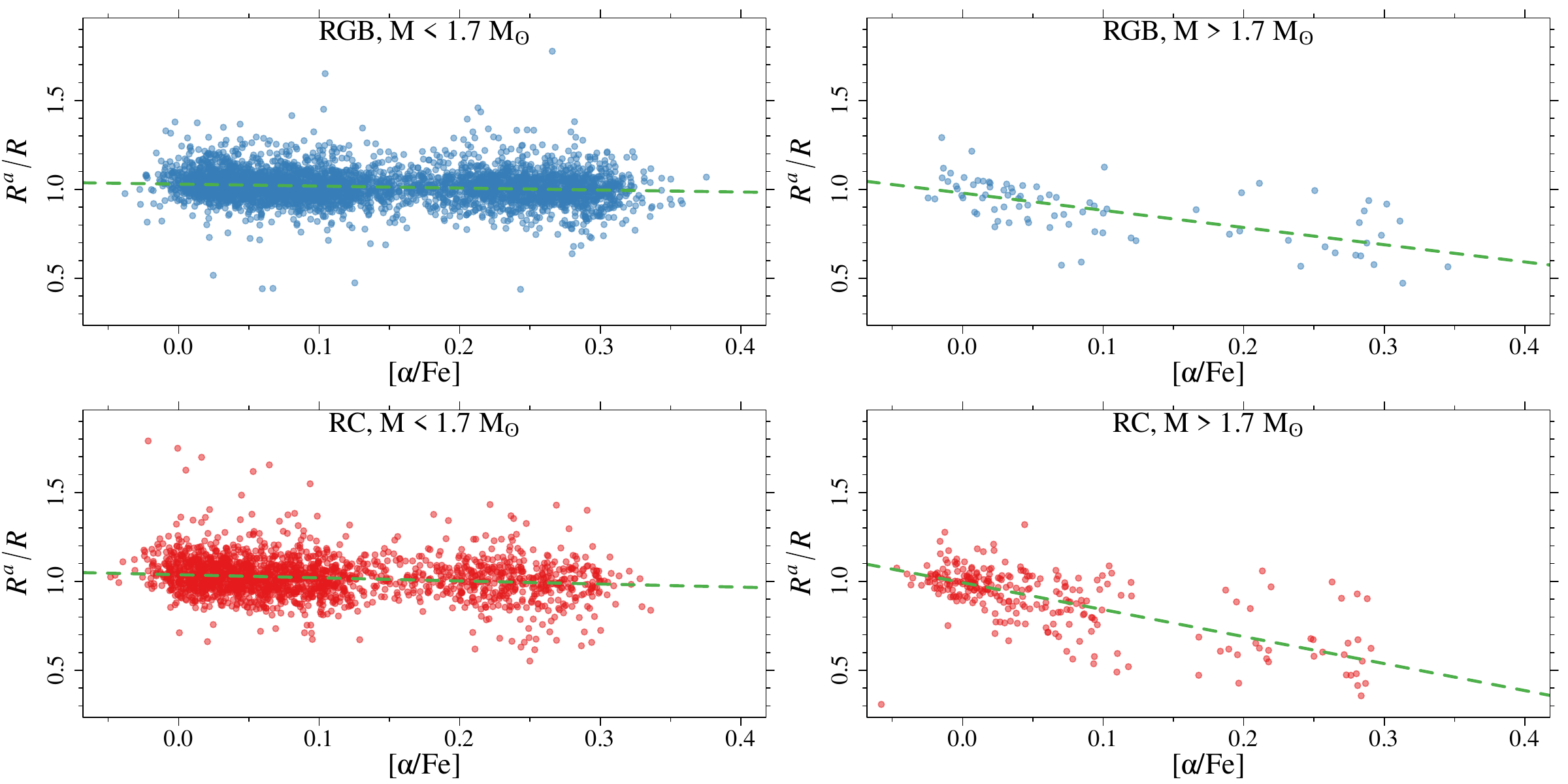}
        \caption{Ratio of the radii from the  \citet{Pietrzynski2019} SBCR and asteroseismology as a function of [$\alpha$/Fe] in different evolutionary phases and mass bins.  {\it Top row}: Radius ratio for RGB stars. The dashed green lines show the robust regression trends. {\it Bottom row}: Same as in the top row but for RC stars.   }
        \label{fig:R-alpha}
\end{figure*}

\begin{table}
\centering
\caption{Robust regression fits of $R^a/R$ as a function of [$\alpha$/Fe], according to the evolutionary phase and the stellar mass.}\label{tab:rlm}
\begin{tabular}{lcc}
  \hline\hline
 & Intercept & Slope \\ 
  \hline
  RGB, $M<1.7\; M_{\sun}$ & $1.030 \pm 0.002$ & $-0.111 \pm 0.012$ \\ 
  RGB, $M>1.7\; M_{\sun}$ & $0.979 \pm  0.018$ & $-0.967 \pm 0.121$ \\ 
  RC, $M<1.7\; M_{\sun}$ & $1.037 \pm 0.003$ & $-0.174 \pm 0.024$ \\ 
  RC, $M>1.7\; M_{\sun}$ & $0.993 \pm 0.008$ & $-1.517 \pm 0.084$ \\ 
   \hline
\end{tabular}
\end{table}

The observed dichotomous trend, most prominent for the RC population, arises from the differing responses of SBCR-estimated and asteroseismic radii to variations in [$\alpha$/Fe] across different mass ranges. Figure~\ref{fig:R-alpha} shows the $R^a/R$ ratio as a function of [$\alpha$/Fe] for RGB and RC stars, further subdivided by mass. A cutoff value of 1.7 $M_{\sun}$ was chosen, but any value up to 2.0 $M_{\sun}$ yields similar results. Using a higher cutoff would significantly reduce the number of stars in the massive RGB bin.
While radius ratios are almost constant in the lower-mass bin, significant trends with [$\alpha$/Fe]  occur for massive stars. 
To further explore the trends, robust linear models\footnote{
Robust linear models are an alternative to least-squares regression when the assumptions about the error distribution are violated. These models can downplay the influence of outliers and provide more accurate estimates in the presence of non-normal errors. Further details can be found, for example, in \citet{venables2002modern} and \citet{Feigelson2012}.} were fitted to data. The results are presented in Table~\ref{tab:rlm} and over-plotted in Fig.~\ref{fig:R-alpha}. The trends are similar between RGB and RC stars, with a steeper dependence on [$\alpha$/Fe] for the more massive RC population. 
These behaviours suggest a theoretical problem in our understanding of SBCR and asteroseismic radius determinations for $\alpha$-enhanced massive stars. However, this question cannot be solved with the available data, and further theoretical research is needed to assess the origin of the discrepancy.
The evident difference between SBCRs and asteroseismic radii for the RC population in Fig.~\ref{fig:R-R2-resid} and the dichotomous trend in Fig.~\ref{fig:R-mass-feh} arise from the superposition of the trends in the different mass bins.

\begin{figure}
        \centering
        \resizebox{\hsize}{!}{\includegraphics{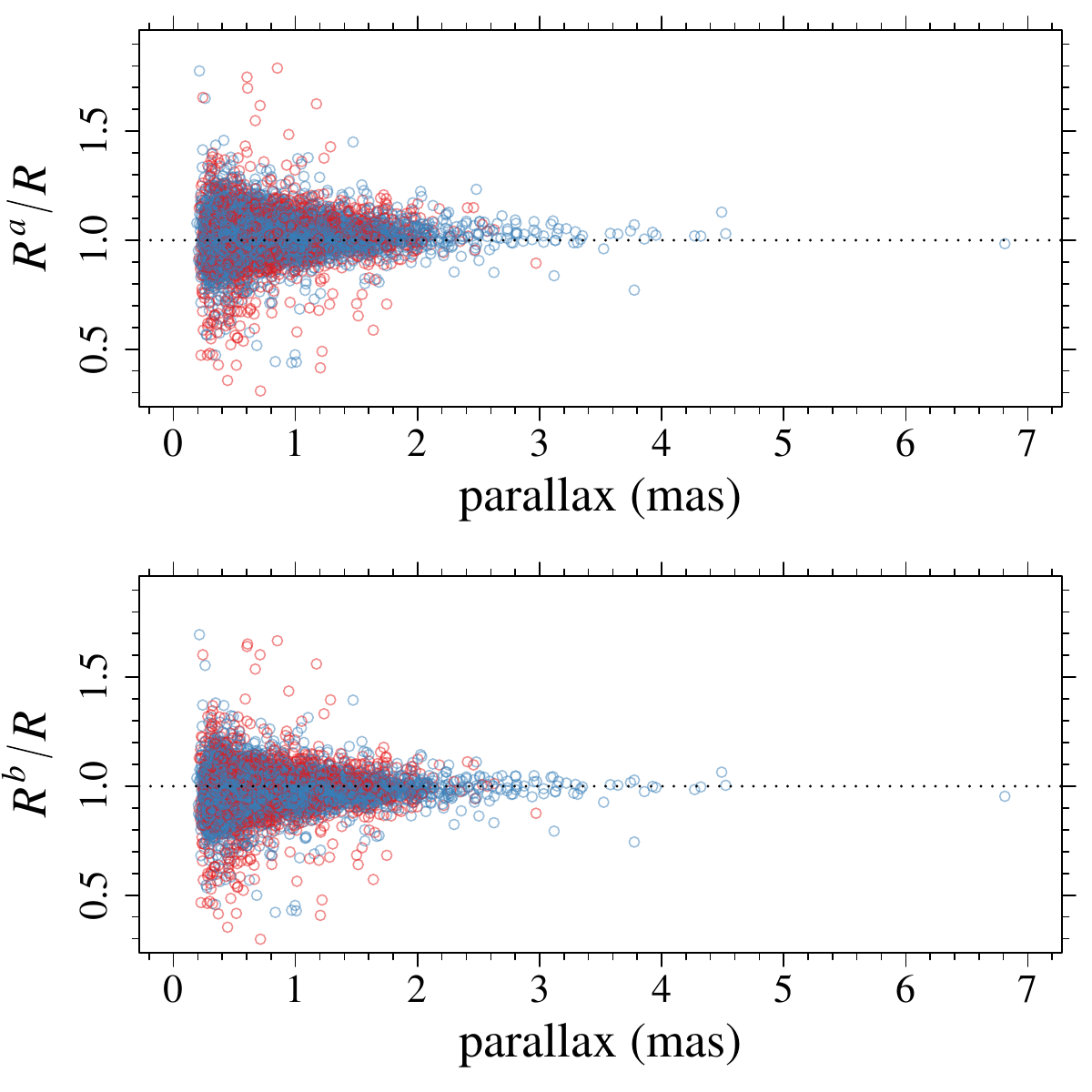}}
        \caption{Ratio of the radii from  SBCRs and from asteroseismology as a function of the {\it Gaia} DR3 parallax.  {\it Top}: Radius ratio for the \citet{Pietrzynski2019} SBCR. {\it Bottom}: Same as in the top panel but for the \citet{Salsi2021} SBCR. Colour codes are as in Fig.~\ref{fig:R-R2-resid}.   }
        \label{fig:R-R2-para}
\end{figure}

As expected, the agreement between the radii from SBCRs and asteroseismology is significantly better for nearby stars, where parallax measurements are most accurate. Figure~\ref{fig:R-R2-para} demonstrates the noticeable decrease in the dispersion of radius ratios as the parallax increases. The standard deviation of the radius ratios for both SBCRs is halved for stars with a parallax greater than 2.5 mas, reaching 6\%. Interestingly, the \citet{Salsi2021} SBCR perfectly agrees with the asteroseismic radii in this range, while the \citet{Pietrzynski2019} method has a positive median overestimation of about 3.0\%.

\section{Conclusions}\label{sec:conclusions}

Leveraging the recently released APO-K2 catalogue \citep{Stasik2024}, which contains astrometric, spectroscopic, and asteroseismic information of more than 7,500 stars in the RGB and RC evolutionary phases, we performed a comparison of radii derived from asteroseismic scaling relations and those from SBCRs combined with {\it Gaia} DR3 parallaxes. We specifically adopted SBCRs from \citet{Pietrzynski2019} and \citet{Salsi2021}. Information about colour excess and magnitude in the $V$ band was obtained by cross-matching the APO-K2 catalogue with TIC v8.2.
The adopted stellar sample is a couple of orders of magnitude larger than those adopted to investigate the agreement between asteroseismic estimates of stellar radii in binary systems \citep[e.g.][]{Gaulme2016, Brogaard2018, Themessl2018}. The sample
encompasses different evolutionary stages, from the early RGB to the RC, and covers a wide metallicity range ([Fe/H] from -1.0 to 0.5 dex). It therefore allowed us to test the agreement
between the different radius estimates as a function of various stellar evolutionary characteristics.

Overall, both SBCRs demonstrated good agreement with the asteroseismic estimates. Since asteroseismic scaling relations and SBCRs rely on very different information to obtain their estimates, this agreement is particularly relevant. 
Comparisons were made using asteroseismic values as a reference, though this choice does not assume them to be unbiased estimators 
\citep[see][and references therein for a discussion of the accuracy and precision of asteroseismic radii]{Hekker2020}.
The SBCR from \citet{Pietrzynski2019} was found to predict a larger radius with respect to scaling relations by about 1.2\% over the whole explored range, while the SBCR from \citet{Salsi2021} was found to be  2.5\% smaller than asteroseismic values. Both SBCRs led to radii significantly smaller than the asteroseismic ones when the latter are larger than about 20 $R_{\sun}$. The two SBCR methods show a similar dispersion of about 10\% in the ratio of radii with asteroseismic estimates.
The agreement with asteroseismic radii is found to depend strongly on the parallax because the dispersion is halved for stars with a parallax greater than 2.5 mas, where \citet{Salsi2021} estimates proved to be in perfect agreement with asteroseismic ones, while \citet{Pietrzynski2019}  radii were overestimated by about 3\% with respect to asteroseismic values.

A trend with metallicity [Fe/H] is found, at a level between  4\% and 6\% per dex. An impact of the metallicity on SBCRs has already been suggested by \citet{Kervella2004} and \citet{Boyajian2012} for main-sequence dwarf stars.
For giant stars, a recent theoretical investigation by \citet{Salsi2022}, adopting MARCS model atmospheres \citep{Gustafsson2008} to compute spectra and obtain the surface brightness of stars, found a slight impact of the metallicity, which was almost negligible at the effective temperature of about 5000 K and increased to 10\% per dex at lower temperatures, corresponding to $V-K \sim 4$.  The detected trend of the radius ratios with metallicity is therefore higher than the theoretical prediction of the SBCRs, suggesting a possible bias in the asteroseismic radii.

We  found a clear trend with the asteroseismic mass. Stellar radii from SBCRs are significantly larger, by about 6\%, than the asteroseismic ones for stars less massive than 0.95 $M_{\sun}$, while no difference was detected in the 1.0 to 2.0 $M_{\sun}$ mass range. This overestimation is correlated with the presence of a significant extension of the helium core with respect to the envelope in the stellar structures.
The most interesting result of the present investigation is that the radius ratio trends with asteroseismic mass show a dichotomous behaviour moving to high mass values, mainly due to the presence of several RC stars with SBCR radii significantly lower with respect to asteroseismology. 
Analysis of this behaviour allowed us to highlight a different response of the asteroseismic scaling relations and SBCRs to [$\alpha$/Fe] for massive stars,  in both the RGB and RC phases, which is reported here for the first time. 
In fact, theoretical investigation by \citet{Salsi2022} did not detect any impact of the mass on their synthetic SBCR, but the tested mass range did not comprise low-mass stars, extending from 2.0 to 5.0 $M_{\sun}$. Moreover no investigation was performed to assess the influence, if any, of $\alpha$ enhancement. Further investigations are encouraged in order to understand these trends because they suggest a theoretical problem in the SBCR and/or asteroseismic radius estimates.

The strong agreement between radii determined by SBCRs and asteroseismology, particularly for RGB stars, is highly encouraging. These two methods differ fundamentally, not only in the observables they rely on but also in terms of observation time, instrumentation requirements, and budget constraints.
In all of these aspects, SBCRs offer significant advantages, opening the possibility for obtaining large samples of asteroseismic-compatible radius estimates with minimal financial investment.

\begin{acknowledgements}
G.V., P.G.P.M. and S.D. acknowledge INFN (Iniziativa specifica TAsP) and support from PRIN MIUR2022 Progetto "CHRONOS" (PI: S. Cassisi) finanziato dall'Unione Europea - Next Generation EU.
\end{acknowledgements}

\bibliographystyle{aa}
\bibliography{biblio}

\end{document}